\documentclass[namedreferences]{SolarPhysics}
\usepackage[optionalrh]{spr-sola-addons} 
\usepackage{graphicx}                    
\usepackage{color}                       
\usepackage{url}                         

\renewcommand{\vec}[1]{ {\mathbf #1} }

\begin{document}
\begin{article}
\begin{opening}

\title{Magnetic Field Structures in a Facular Region Observed by THEMIS and {\it Hinode}}

\author{Y.~\surname{Guo}$^{1,2}$\sep
        B.~\surname{Schmieder}$^{1}$\sep
        V.~\surname{Bommier}$^{3}$\sep
        S.~\surname{Gosain}$^{4}$
       }

\runningauthor{Guo et al.}
\runningtitle{Magnetic field structures
in a facular region}

\institute{$^{1}$ Observatoire de Paris, Section de Meudon, LESIA, 92195 Meudon Principal Cedex, France\\
                   email: \url{yang.guo@obspm.fr}\\
           $^{2}$ Department of Astronomy, Nanjing University, Nanjing 210093, China\\
           $^{3}$ Observatoire de Paris, Section de Meudon, LERMA, 92195 Meudon Principal Cedex, France\\
           $^{4}$ Udaipur Solar Observatory, P. Box 198, Dewali, Udaipur 313001, India
          }

\begin{abstract}
The main objective of this paper is to build and compare vector
magnetic maps obtained by two spectral polarimeters, {\it i.e.}
THEMIS/MTR and {\it Hinode} SOT/SP, using two inversion codes
(UNNOFIT and MELANIE) based on the Milne--Eddington solar
atmosphere model. To this end, we used observations of a facular
region within active region NOAA 10996 on 23 May 2008, and found
consistent results concerning the field strength, azimuth and
inclination distributions. Because SOT/SP is free from the seeing
effect and has better spatial resolution, we were able to resolve
small magnetic polarities with sizes of $1''$ to $2''$, and we could
detect strong horizontal magnetic fields, which converge or
diverge in negative or positive facular polarities. These findings
support models which suggest the existence of small vertical flux tube
bundles in faculae. A new method is proposed to get the relative
formation heights of the multi-lines observed by MTR assuming the
validity of a flux tube model for the faculae. We found that the
Fe {\sc \romannumeral 1} 6302.5 \AA \ line forms at a greater
atmospheric height than the Fe {\sc \romannumeral 1} 5250.2 \AA \
line.
\end{abstract}

\keywords{Active Regions, Structure; Magnetic fields, Photosphere;
Polarization, Optical}

\end{opening}

\section{Introduction}\label{s-intro}

Faculae are bright plages observed in H$\alpha$ or photospheric
lines. They are considered as magnetized regions constituted of a
bundle of thin vertical flux tubes with a magnetic field strength of
thousands of gauss and a size of tens to hundreds of kilometers
\cite{Zwaan1987}. Similar values for field strength and size of the
flux tubes have been found by analyzing the ratio of Stokes $V$
signals in the Fe {\sc \romannumeral 1} 5250.2 \AA, Fe {\sc
\romannumeral 1} 5247.1 \AA, and Fe {\sc \romannumeral 1} 5232.9
\AA ~lines, for a spatially unresolved photospheric network
\cite{Stenflo1973}. The flux tubes are located at the boundaries
of granules and bundled by the convection flow. Since the pressure
of the solar atmosphere decreases with height, the tubes expand and
the magnetic field strengths decrease to compensate the pressure
decrease. The network field expands rapidly with height and forms
a canopy above the internetwork region. Similarly, the faculae
would be bundles of flux tubes expanding with height.

In order to derive the three-dimensional (3D) magnetic structure
of these features, we require multi-line spectral polarimetry. On
the other hand, to resolve the fine structure, we require high
spatial resolution. The French--Italian ground-based solar
telescope THEMIS (T\'elescope H\'eliographique pour l'Etude du
Magn\'etisme et des Instabilit\'es Solaires) on the Canary Islands
\cite{Lopez2000,Bommier2007} and the Japanese space borne
satellite {\it Hinode} \cite{Kosugi2007} provide this ability.
THEMIS is designed to be free from the instrumental polarization,
since the polarization analyzer is located at the primary focus.
It is also characterized by the multi-line spectroscopy capability
in the MTR (Multi-Raies) grid mode that ensures the best
co-spatiality. The beam exchange technique of MTR increases the
polarimetric accuracy as the flat field errors are minimized.
Solar Optical Telescope (SOT) aboard the {\it Hinode} spacecraft
is not affected by the Earth's atmospheric seeing effect. It
observes diffraction limited images with $0.2''$ -- $0.3''$
resolution by its 50 cm aperture optical telescope. The Spectral
Polarimeter (SP) behind SOT observes the full Stokes profiles of
two magnetically sensitive lines Fe {\sc \romannumeral 1} 6301.5
\AA \ and Fe {\sc \romannumeral 1} 6302.5 \AA \
(\opencite{Ichimoto2008}; \opencite{Shimizu2008};
\opencite{Suematsu2008}; \opencite{Tsuneta2008}). With its high
performance, breakthroughs in different topics related to
photospheric fields have been obtained by SOT: the detection of
horizontal fields over granules and the hidden turbulent magnetic
flux \cite{Lites2007}, the birth of small flux tubes with
kilo-gauss field strength \cite{Nagata2008}, and the discovery of
transient horizontal magnetic fields in quiet sun
\cite{Ishikawa2008}.

The aim of this paper is: (i)  to demonstrate that the two
spectral polarimeters THEMIS/MTR and {\it Hinode} SOT/SP, and the
two Stokes profile inversion codes give consistent results, and
(ii) to study the magnetic structure of faculae. We observe a
facular region, build and compare vector magnetic fields using the
full Stokes profiles of multi spectral lines observed by two
spectral polarimeters (MTR and SOT/SP) and fitted by two inversion
codes. We find concentrations of magnetic flux with converging and
diverging transverse field vectors in high angular resolution
magnetograms observed by SOT/SP. Finally, we use a flux tube model
to get a qualitative view on the formation heights of two Fe {\sc
\romannumeral 1} lines. The multi-line capability of THEMIS/MTR
and the high angular resolution of SOT/SP complement each other in
this study. The description of the observations and the methods of
data analysis are given in Section \ref{s-obser}. The results are
presented in Section \ref{s-resul}. We discuss the magnetic field
gradient according to the flux tube model in Section
\ref{s-magalt}. The conclusions are drawn in Section
\ref{s-discu}.

\section{Observations and Data Analysis}\label{s-obser}

\subsection{Instruments and Observations}
\begin{figure}
\centerline{\includegraphics[width=0.8\textwidth,viewport=20 50
508 488, clip]{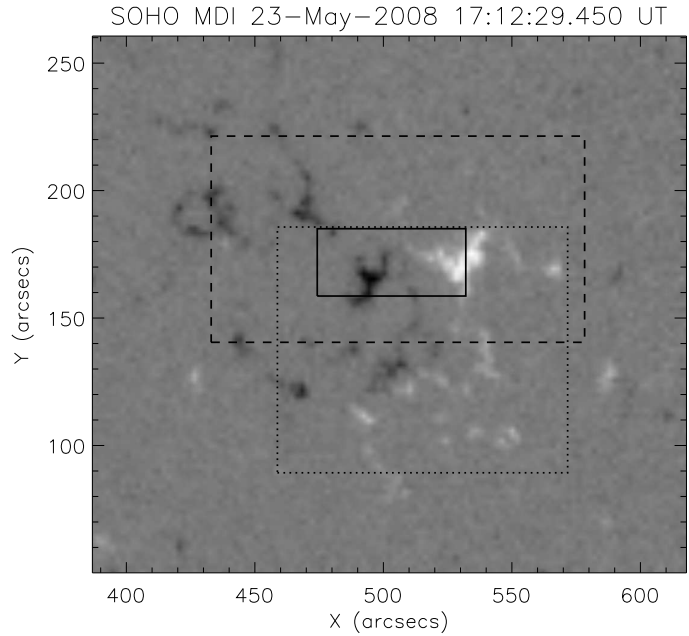}} \caption{The full fields of view of
SOT/SP (dashed rectangle) and THEMIS/MTR (dotted rectangle). The
solid rectangle denotes the region that we select to study the
magnetic field in detail. The background is part of the full disk
96 min magnetogram observed by SOHO/MDI.}\label{fig01}
\end{figure}

The faculae in the active region NOAA 10996 (N10$^\circ$
W32$^\circ$) were simultaneously observed by THEMIS and {\it
Hinode}. THEMIS/MTR scanned the region from 16:44--17:38 UT and
{\it Hinode} SOT/SP from 14:30--15:01 UT, on 23 May 2008. The
heliocentric angle for the observed region was $\approx 34^\circ$.
We plot the full field of view of SOT/SP and THEMIS/MTR in Figure
\ref{fig01}. Both instruments scan the solar surface from east to
west. THEMIS/MTR recorded the two beams on the same CCD camera
using a grid in order to avoid problems of co-spatiality. The grid
had empty space ($15.5''$) between the bars, along the slit
($\approx 120''$). The grid bars were covered with three $8''$
steps along the slit (which was north--south). The next step was to
move the telescope towards the west with a step size of $0.8''$.
The scan of the region ($110''\times104''$) was obtained within
about one hour. The characteristics of THEMIS/MTR are given in
Table \ref{tbl11}.

The three photospheric lines observed with MTR were selected
because they are all normal Zeeman triplet lines (formed by
transition between energy levels with $J=0$ and $J=1$). Moreover,
their Land{\'e} $g$ factors are rather large, 3.0, 2.5
and 2.0 for Fe {\sc \romannumeral 1} 5250.2 \AA, Fe {\sc
\romannumeral 1} 6302.5 \AA,  and Ca {\sc \romannumeral 1} 6102.7
\AA, respectively. For both reasons, these lines in the visible
spectrum have good magnetic sensitivity. They are also well
adapted to the UNNOFIT inversion, which can treat only normal
Zeeman triplet lines.  For the other lines we have UNNOFIT2
\cite{Bommier2007}, but generally their Zeeman sensitivity is
lower and the inversion fails to converge in pixels with poor
signal-to-noise ratio.

The fast map mode of SOT/SP provides us with a relatively high
cadence of 3.8 s per position of the slit, and a spatial
resolution of $\approx 0.3''$ per pixel after binning on flight.
The width of the slit was $0.16''$, and the length was reduced to
$164''/2$ for this study. The scan of the region
($145''\times82''$) was obtained within 30 minutes. We selected a
central part of the active region, where data were observed with
good quality both by SOT/SP and MTR (solid rectangle in
Figure~\ref{fig01}), to perform a detailed analysis.

\subsection{Inversion Procedures and Bisector Method}
How can we extract the vector magnetic field from the full Stokes
profiles? It needs two steps. The forward problem answers how the
line forms for given magnetic fields in the photosphere of the Sun,
by solving the radiative transfer equations for polarized
radiation \cite{Unno1956,Rachkovsky1962,Jefferies1989}. The
inversion problem extracts the magnetic field and other physical
parameters by fitting the observed line profiles with the
theoretical line profiles obtained in the first step
\cite{Landolfi1982,Skumanich1987}.

All the raw spectra of THEMIS are calibrated by spectral
destretching, dark current subtraction and flat field correction
\cite{Bommier2002a,Bommier2002b} before polarimetric analysis and
inversion. In Table \ref{tbl11}, we list the observational
characteristics of five of the major lines that we have used
within the wavelength range of MTR, and of the Fe {\sc
\romannumeral 1} 6302.5 \AA \ line of SOT/SP. The $I, Q, U,$ and
$V$ profiles of SOT/SP are obtained by adding and subtracting the
raw spectra after calibration with the  standard data analysis
packages for SOT/SP in Solar Software (SSW) \cite{Ichimoto2008}.

We use the UNNOFIT inversion code \cite{Landolfi1982,Bommier2007}
to fit the line profiles of Ca {\sc \romannumeral 1} 6102.7 \AA,
Fe {\sc \romannumeral 1} 6302.5 \AA, and Fe {\sc \romannumeral 1}
5250.2 \AA \ observed by THEMIS/MTR. The data obtained by SOT/SP
are fitted by two inversion codes, {\it i.e.} by UNNOFIT and by
the code developed by {\it Hinode} team based on MELANIE
(Milne--Eddington Line Analysis using an Inversion Engine;
\opencite{Socas-Navarro2001}). UNNOFIT and MELANIE use the
Levenberg--Marquardt algorithm to achieve the least $\chi^2$
fitting of the observed profiles by the  model profiles. The model
profiles are given by the Unno--Rachkovsky solution based on the
Milne--Eddington approximation for the solar atmosphere. Including
the magneto-optical and damping effects, all parameters needed to
synthesize the theoretical Stokes profiles are listed in Table
\ref{tbl12}. The damping parameter $\gamma$ and damping constant
$a$ are related by $a=\gamma/(4\pi\Delta\lambda_{\mathrm D})$. The
parameter of the source function $b$ is defined by $b=\mu
B_1/B_0$, where $\mu$ is the cosine of the angle between the
normal to the solar surface and line-of-sight (LOS), and
$B(\tau)=B_0+B_1\tau$, where $\tau$ is the optical depth.
\inlinecite{Bommier2007} introduced the filling factor in UNNOFIT
as follows:
\begin{equation}
\left\{
\begin{array}{l}
I=(1-\alpha)I_\mathrm{q}+\alpha I_{\mathrm m} \\
Q=\alpha Q_{\mathrm m} \\
U=\alpha U_{\mathrm m} \\
V=\alpha V_{\mathrm m}
\end{array} \right..
\end{equation}
$I_\mathrm{m}$, $Q_{\mathrm m}$, $U_{\mathrm m}$, and $V_{\mathrm
m}$ are the Stokes parameters in magnetic fields. $I_\mathrm{q}$
is the intensity of the quiet sun. They have the same physical
conditions except for the presence or absence of magnetic fields. So
the profiles of $I_\mathrm{q}$ in UNNOFIT are calculated using the
radiative transfer equations with the same parameters as used for
$I_{\mathrm m}$, while ignoring the magnetic fields. Note that
$I_\mathrm{q}$ in MELANIE is given by the average intensity
profiles within the whole field of view excluding active regions,
and it is usually recognized as the stray light
\cite{Skumanich1987}, whose fraction $f$ is relative to the
filling factor $\alpha$ as $\alpha=1-f$. All the magnetic field
maps that are plotted in the figures of this paper have been
multiplied by the filling factors. They are $\alpha B$ for UNNOFIT
inversions and $(1-f)B$ for MELANIE inversions. These local
averaged magnetic fields are the only determined quantities, since
the intrinsic field strengths and filling factors cannot be
determined separately as shown by \inlinecite{Bommier2007}. From
now on, we will use $B$ instead of $\alpha B$ to represent the
local average field strength for concision. But we should keep in
mind that all field strengths and field components, such as the
LOS magnetic fields, have been multiplied by the filling factors.

The magnetic fields of the Na D$_1$ line profiles observed by THEMIS/MTR
are calculated by the formula \cite{Semel1967} as
follows:
\begin{equation}
\delta\lambda=4.67 \times 10^{-13} g_\mathrm{eff} \lambda^2
B_\mathrm{LOS} \ ,\label{eqn12}
\end{equation}
where $g_\mathrm{eff}=1.33$ for the Na D$_1$ line, and the units
of wavelength and magnetic field are \AA \ and gauss,
respectively. The formula shown in Equation (\ref{eqn12}) is
different from the weak field approximation, where the LOS
magnetic field is calculated by $V \propto B_\mathrm{LOS}
\frac{\partial I}{\partial \lambda}$. Equation (\ref{eqn12})
remains valid even for strong fields \cite{Bommier2005}. We
determined the Zeeman shift $\delta\lambda$ by a bisector method
using the middle point of a chord with a length of $2\Delta\lambda$,
which connects the equal intensity points of a line profile. The
wavelength difference of the middle points of the $I+V$ and $I-V$
profiles is the Zeeman shift $\delta\lambda$ \cite{Berlicki2006},
and $\Delta\lambda$ represents the wavelength difference from the line
core approximately. We obtain the magnetic fields for
$\Delta\lambda$ = 0.08 \AA, 0.16 \AA, 0.24 \AA, and 0.32 \AA \ in
the Na D$_1$ line.

\subsection{Co-alignment}\label{s-align}
In order to compare the magnetic field characteristics, we have to
co-align the images in different wavelength bands and by different
instruments. The observations in different wavelength bands of MTR
have offsets between each other. The pixel sizes for these
wavelength bands also differ slightly, which requires an
interpolation to the same pixel resolution before co-alignment.
Also, the instruments observe only partial regions of the solar
disk; hence pointing errors cannot be avoided. We use the magnetic
map of the Michelson Doppler Imager (MDI; \opencite{Scherrer1995})
as the basic reference frame, since the full-disk observation
ensures the accuracy of the MDI coordinates by comparing the
position of solar limbs \cite{Guo2009}. Firstly, we make a coarse
alignment of the MTR and SOT/SP magnetic maps referred to the MDI
magnetic map that has the closest observation time. The so-called
feature identification method is employed, where several points
sharing common features on the two images are selected
consecutively to determine the shifts between them. Next, the
magnetic maps of MTR are interpolated to a common spatial
resolution, which is that of Fe {\sc \romannumeral 1} 6302.5 \AA \
here. The magnetic map of SOT/SP is differentially rotated to the
middle observation time of the MTR scanning and interpolated.
Finally, the fine alignment is done by correlating each image with
the Fe {\sc \romannumeral 1} 6302.5 \AA \ magnetic map, and
finding out the position where two images have the largest
correlation coefficient.

\begin{figure}
\centerline{\includegraphics[width=0.8\textwidth]{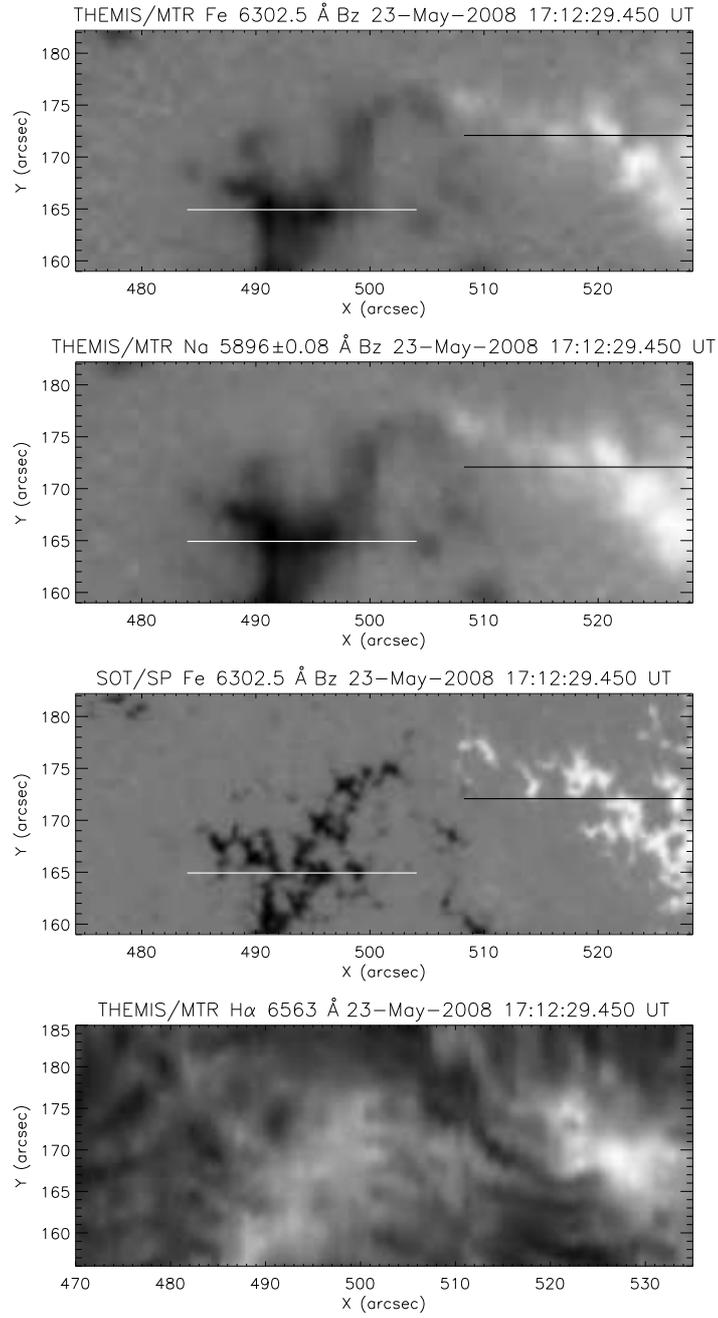}}
\caption{LOS magnetic fields of THEMIS/MTR Fe
{\sc \romannumeral 1} 6302.5 \AA, Na D$_1$ 5896$\pm$0.8 \AA, and
SOT/SP Fe {\sc \romannumeral 1} 6302.5 \AA \ in the solid rectangle
region as shown in Figure \ref{fig01}. The H$\alpha$ intensity map
is shown in the bottom panel. Some slices are drawn and denoted by the
solid lines on the magnetic field images, which will be used in Figures
\ref{fig06} and \ref{fig08}.}\label{fig02}
\end{figure}

We present in Figure \ref{fig02} some of the co-aligned maps of
the LOS magnetic fields in the selected field of view, shown by
the solid rectangle in Figure \ref{fig01}. For MTR, magnetic
fields have been calculated for the Ca {\sc \romannumeral 1}
6102.7 \AA, Fe {\sc \romannumeral 1} 6302.5 \AA, Fe {\sc
\romannumeral 1} 5250.2 \AA, and Na D$_1$ 5896 \AA \ lines, and
for the SOT/SP Fe {\sc \romannumeral 1} 6302.5 \AA \ line. For the
Na D$_1$ line we have calculated four magnetic maps at
$\Delta\lambda$ = 0.08 \AA, 0.16 \AA, 0.24 \AA, and 0.32 \AA. Only
one Na D$_1$ magnetic map for $\Delta\lambda = 0.08$ \AA \ is
shown Figure \ref{fig02} as an example. We draw two cuts in each
magnetic field map in order to perform a detailed comparison of
the field distributions in the subsequent sections.

\subsection{180$^\circ$ Ambiguity Removal}
The transverse component of the magnetic field vector has an
intrinsic 180$^\circ$ ambiguity. We have to remove this ambiguity
before we can transform the magnetic field to the local
heliographic coordinates for a better understanding of the
magnetic field structure. \inlinecite{Metcalf2006} reviewed
different existing methods that aim to remove the 180$^\circ$
ambiguity. A new method based on the fields measured at two
different heights and the divergence-free nature of the magnetic
field has been recently proposed by \inlinecite{Crouch2009} and
is currently under development.

In this work we adopt the method of non-potential magnetic field
calculation, developed by \inlinecite{Georgoulis2005} and
\inlinecite{Metcalf2006}. This method iteratively compares the
observed magnetic field vectors with the model field $\vec B =\vec
B_\mathrm{p} +\vec B_\mathrm{c}$ in the heliographic coordinates
system, where the potential component $\vec B_\mathrm{p}$ is
calculated by potential field extrapolation and the non-potential
component $\vec B_\mathrm{c}$ is calculated by the direct and
inverse Fourier transforms of the vertical electric current
density $J_z$ \cite{Chae2001,Georgoulis2005}. We use the improved
version of the non-potential magnetic field calculation method
\cite{Metcalf2006}, where the vertical electric current density
$J_z$ is updated in every iterative step based on the intermediate
ambiguity solution $\vec B$. $J_z$ is calculated by means of
Amp\`ere's law. The field of view for the potential and
non-potential magnetic field is padded with zeros to allow for a more
accurate implementation of the Fourier transform.

\subsection{MTR H$\alpha$ Observation}
We plot the intensity map of the facular region at the line center
of H$\alpha$ 6563 \AA \ in the bottom panel of Figure \ref{fig02}.
The H$\alpha$ map is not precisely aligned with the magnetic field
map, but the alignment is sufficient for a qualitative comparison.
The comparison shows that dark filaments are located between
positive and negative magnetic polarities. The H$\alpha$ intensity
increases in the regions of magnetic field concentration. This can
also be found in the {\it Hinode} SOT/BFI (Broadband Filter
Imager) Ca {\sc \romannumeral 2} H line intensity map. It is still
not fully understood why faculae are brighter than a normal quiet
sun region. One possible explanation is that the optical thickness
is smaller in magnetic tubes than in the external photosphere.
This effect allows us to see deeper and hotter layers
\cite{Strous1994}. \inlinecite{Keller2004} have performed a 3D
simulation of non-grey radiative magneto-convection to explain the
origin of this phenomenon. \inlinecite{Hasan2008} suggest that the
bright features are heated by the dissipation of short period
magneto-acoustic waves in magnetic flux tubes.

\section{Results}\label{s-resul}

\subsection{Magnetic Field Maps and Histograms}

\begin{figure}
\centerline{\includegraphics[width=0.8\textwidth]{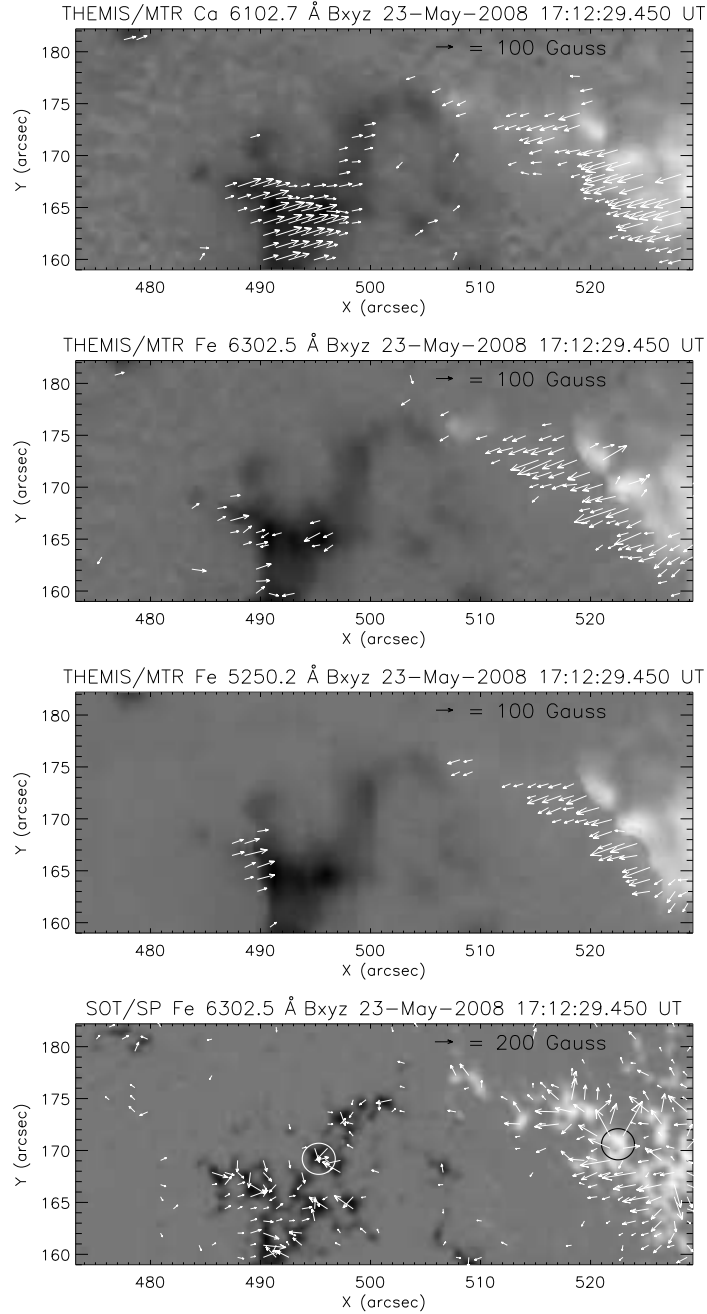}}
\caption{The vector magnetic fields of THEMIS/MTR Ca {\sc
\romannumeral 1} 6102.7 \AA, Fe {\sc \romannumeral 1} 6302.5 \AA,
Fe {\sc \romannumeral 1} 5250.2 \AA, and SOT/SP Fe {\sc
\romannumeral 1} 6302.5 \AA \ in the solid rectangle region as
shown in Figure \ref{fig01}. The 180$^\circ$ ambiguity has been
removed by the non-potential magnetic field calculation method and
the fields have been transformed into the heliographic
coordinates. Backgrounds are the vertical fields, white (black) is
positive (negative). Arrows are the horizontal fields. The
threshold of the plotted arrows is 50 gauss. The circles in the
bottom panel denote the regions of divergence and convergence of
the arrows.}\label{fig03}
\end{figure}

Magnetic maps of MTR Ca {\sc \romannumeral 1} 6102.7 \AA, Fe {\sc
\romannumeral 1} 6302.5 \AA, Fe {\sc \romannumeral 1} 5250.2 \AA \
and SOT/SP Fe {\sc \romannumeral 1} 6302.5 \AA \ are shown in
Figure \ref{fig03}. All magnetic field parameters have been
transformed into a local solar coordinate system. The arrows are
plotted for the transverse fields with strength larger than 50
gauss, which is the limit of the transverse field accuracy
\cite{Ichimoto2008}. The results obtained from MTR show that in
the regions of strong concentrations, the magnetic fields are
almost vertical. The horizontal components are typically less than
50 gauss; therefore, only a few arrows appear in these regions. In
the central part of the facular region the arrows could correspond
to the footpoints of H$\alpha$ fibrils that appear in the bottom
panel of Figure \ref{fig02} between the two bright regions.

The fields obtained by SOT/SP show more detailed features. The
horizontal vectors at the inner edges of the strong field
concentration regions also indicate that this region could
correspond to footpoints of the H$\alpha$ fibrils. In large
concentrations we can detect many small cells with horizontal
fields pointing towards and away from the center of negative and
positive polarities, respectively; see the circles in the bottom
panel of Figure \ref{fig03}. This indicates the existence of
expanding flux bundles rooted in these cells, supporting the
canopy model.

\begin{figure}
\centerline{\includegraphics[width=1\textwidth]{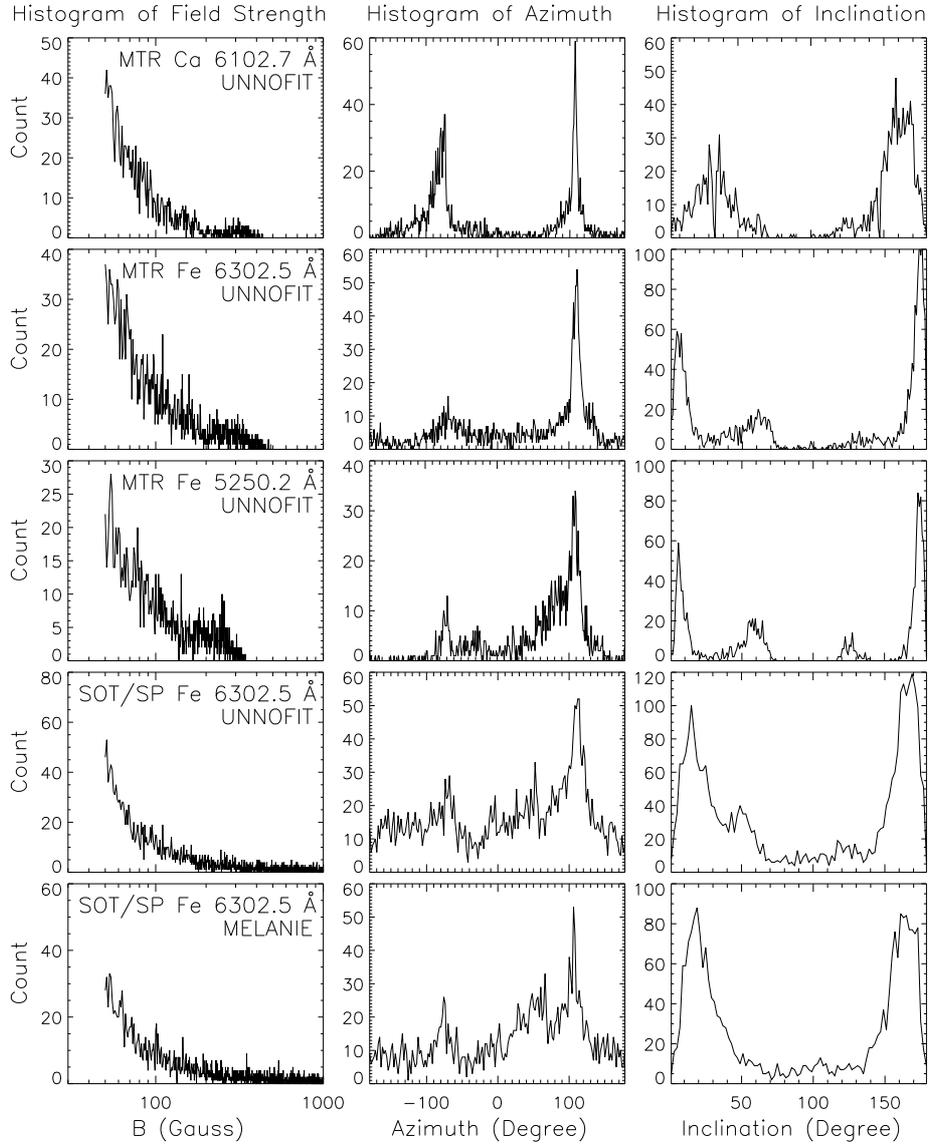}}
\caption{Histograms of the product of field strengths and filling
factors, the azimuth, and inclination angles in the regions as
shown in Figure \ref{fig02} in the heliographic coordinates
system. Only the fields where the field strength is greater than 50
gauss are included. The azimuth angles are measured
counterclockwise from the north direction viewed towards the Sun
and the 180$^\circ$ ambiguity has been removed. The inclination
angles are referred to the vertical direction of the solar
surface. $0^\circ$ is away from the Sun center and $180^\circ$ is
towards the Sun center.}\label{fig04}
\end{figure}

The histograms of the field strength, azimuth, and inclination
angles that have been transformed into the heliographic
coordinates after the $180^\circ$ ambiguity removal are shown in
Figure \ref{fig04}. There are two peaks that appear in each of the
histogram of the azimuth angle. The plots are in the range of
$-180^\circ$ to $180^\circ$, since the ambiguity has been removed.
The histograms of the inclination angles obtained from MTR Fe {\sc
\romannumeral 1} 6302.5 \AA \ and Fe {\sc \romannumeral 1} 5250.2
\AA \ clearly show that the magnetic fields are mainly vertical in
the local solar surface plane, and the inclination histograms of
SOT/SP Fe {\sc \romannumeral 1} 6302.5 \AA \ show departures from
the vertical direction. This is due to the expansion of the flux
tube and the high spatial resolution of SOT/SP. MTR observes the
average value indicating mostly the vertical component.

The inclination angles measured by the THEMIS/MTR Fe {\sc
\romannumeral 1} lines are more vertical  than that measured by
the Ca {\sc \romannumeral 1} 6102.7 \AA \ line. There are several
possible explanations for this difference. First of all, the
uncertainties of the angle measurements may account for it.
\inlinecite{Bommier2007} found that azimuth and inclination angles
can only be determined within the accuracy of $10^\circ$ to
$20^\circ$, due to the polarimetric accuracy of THEMIS/MTR.
Secondly, the magnetic fields obtained by different lines are
formed at different altitudes and the Ca {\sc \romannumeral 1}
line could be formed at greater atmospheric heights than the other
lines, where the magnetic field is more inclined. Finally, the Ca
{\sc \romannumeral 1} 6102.7 \AA \ line may not be formed in local
thermodynamical equilibrium (LTE), whereas UNNOFIT is based on the
Milne--Eddington model of the atmosphere which assumes LTE. In that
case the magnetic fields obtained from the inversion of the Ca
{\sc \romannumeral 1} 6102.7 \AA \ line would have large
uncertainties.

\subsection{Comparison of Magnetic Fields Observed by THEMIS/MTR and SOT/SP}\label{s-fine}

\begin{figure}
\centerline{\includegraphics[width=1\textwidth]{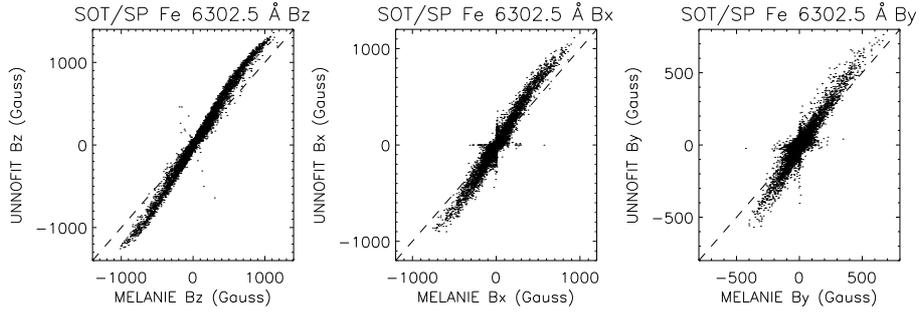}}
\caption{The scatter plots show the comparisons of $B_z$, $B_x$,
and $B_y$ obtained by UNNOFIT and MELANIE inversions and SOT/SP
observation in the field of view of the solid rectangle shown in
Figure \ref{fig01}.}\label{fig05}
\end{figure}

{\it Hinode} observes seeing-free images with better quality and
provides finer structures in this facular region than ground-based
instruments. It makes sense to compare the data from {\it Hinode}
SOT/SP and THEMIS/MTR. In order to make the comparison reliable,
we treated the Stokes profiles of SOT/SP with both UNNOFIT and
MELANIE. The comparison results are shown in Figure \ref{fig05}.
The magnetic fields obtained by UNNOFIT and MELANIE inversion
codes are found coherent, especially for LOS components. Yet,
the absolute values obtained by UNNOFIT inversion code are
generally larger than that by MELANIE inversion code, which can be
explained by the different free parameters considered in the two
inversion codes, especially the different methods of the
computation of the filling factors. Only a few points violate the
positive correlation relationship, and most of them appear in the
weak magnetic field regions.

\begin{figure}
\centerline{\includegraphics[width=1\textwidth]{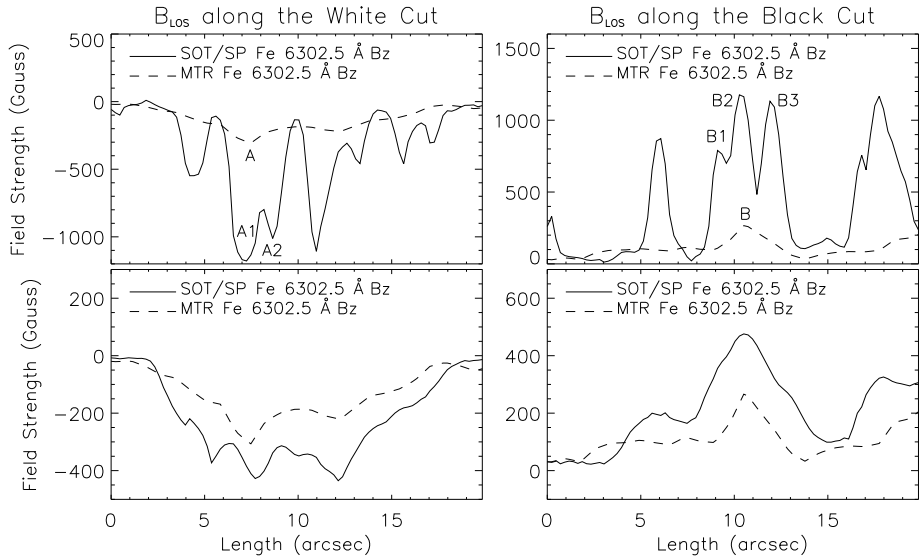}}
\caption{LOS magnetic field distribution along the white cut (left
column) and along the black cut (right column) observed by SOT/SP
(solid line) and MTR (dashed line). Top panels display the fields
obtained by the inversion of the original SOT/SP observed
profiles, while in the bottom row the Stokes profiles were
convolved by a $3''\times 3''$ 2D boxcar function at each
wavelength before inversion. See text for details. }\label{fig06}
\end{figure}

The structures are clearly shown in the top row of Figure
\ref{fig06}, where the LOS magnetic fields obtained by SOT/SP and
MTR Fe {\sc \romannumeral 1} 6302.5 \AA \ are plotted along two
selected cuts. The absolute value of SOT/SP LOS magnetic fields
reaches about 1200 gauss, while the one of MTR is about 300 gauss.
The sizes of the structures are about $1''$ to $2''$ with SOT/SP
if we define them as the full-width at half-maximum of the
absolute value of the field strength variation along a cut, which
are comparable to previous direct measurement
\cite{Muller1994,DaraKoutchmy1983}. These values should correspond
to the sizes of the bundles of flux tubes. The sizes determined by the
observation of MTR in the same region are larger. The magnetic
fields observed by SOT/SP are obtained by the MELANIE inversion. We
have already shown in Figure \ref{fig05} that MELANIE and UNNOFIT
give consistent inversion results. However, considering that
MELANIE has been optimized for SOT/SP data by the {\it Hinode}
team, we still use the MELANIE inversion result for the SOT/SP
data to be compared with the UNNOFIT inversion result of MTR data.
The time difference between the two observations is about 2 hours.
We have checked the evolution of this region by four sets of
magnetograms taken on the same day, and we found no evidence for a
decreasing trend in the field strengths.

The difference of field strengths can be understood by the
difference of pixel sizes ($0.3''$ for SOT/SP and $0.8''$ for
THEMIS/MTR),and by considering the seeing effects and the
photometric accuracy. THEMIS observes on the ground, where the
atmosphere of the Earth blurs, distorts the observed images, and
decreases the spatial resolution. We adopt a convolution process
to simulate the seeing effect on the Stokes profiles observed by
SOT/SP. Each Stokes profile is reconstructed after a convolution
of each wavelength image by a $3''\times 3''$ two-dimensional (2D)
boxcar function sampled at the SOT/SP pixel size. The photometric
accuracy of MTR is $\approx 3\times10^{-3}$ \cite{Lopez2000} and
it is $\approx 10^{-3}$ for SOT/SP \cite{Ichimoto2008}. We add a
random noise with a normal distribution in the field of view and
with a standard deviation of $3I_\mathrm{c}\times10^{-3}$ to each
Stokes profile of SOT/SP at each wavelength. $I_\mathrm{c}$ is the
continuum intensity. Then the MELANIE inversion is applied. The
distributions of the LOS magnetic fields on the selected cuts are
shown in the bottom row of Figure \ref{fig06}. Both the field
strength and the size of the magnetic concentration feature of the
degenerate SOT/SP LOS magnetic field map are comparable to what
MTR observed.

\subsection{Formation Heights of Ca {\sc \romannumeral 1} and Fe {\sc \romannumeral 1} Lines}

Some simulated formation heights of Fe {\sc \romannumeral 1} lines
are listed in Table \ref{tbl13}. The heights are measured from
$\tau_{5000} = 1$, and $\tau_{5000}$ is the optical depth at the
wavelength of 5000 \AA. All the calculations adopt the LTE
assumption, but with different atmosphere models.
\inlinecite{Bommier2007} employed the quiet sun photospheric
reference model \cite{Maltby1986}, extrapolated downwards beyond
$-70$ km to $-450$ km below the $\tau_{5000} = 1$ level. Above
$-70$ km, this model is very similar to the quiet sun VAL-C model
\cite{Vernazza1981}. \inlinecite{Bruls1991} adopted the VAL-C
model, and \inlinecite{Sheminova1998} assumed the
Holweger--M\"uller model \cite{Holweger1974}. The formation heights
show some discrepancies. The calculation of these heights depends
on the choice of the physical parameters, such as the temperature,
magnetic field, and microturbulent velocity. We would need the
formation altitudes of lines by using the response function to
magnetic field perturbations (Vitas, private communication). It
is not in the scope of this paper. However, we propose another way
to derive the relative formation altitudes of the lines by using
the assumption of the flux tube model in the faculae.

\begin{figure}
\centerline{\includegraphics[width=1\textwidth]{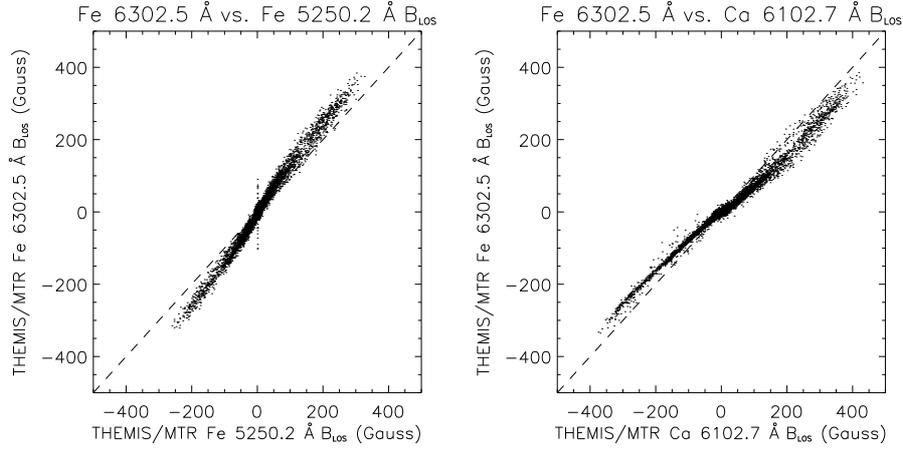}}
\caption{Scatter plots of the LOS magnetic fields obtained by
THEMIS/MTR Fe 6302.5 \AA \ vs. Fe 5250.2 \AA \ (left panel) and by
MTR Fe 6302.5 \AA \ vs. Ca 6102.7 \AA \ (right panel). The data
points are in the field of view of the solid rectangle shown in
Figure \ref{fig01}.}\label{fig07}
\end{figure}

In order to have a quantitative comparison of the LOS magnetic
fields obtained by the different lines of THEMIS/MTR, we produce
scatter plots, shown in Figure \ref{fig07}. The plots show that
for each pixel the LOS magnetic fields obey the following
relation, except for some weak field pixels that we will not
consider, because they are not in facular concentrations:
\begin{equation}
|B_\mathrm{LOS,6102.7}| > |B_\mathrm{LOS,6302.5}| >
|B_\mathrm{LOS,5250.2}| \ .\label{eqn13}
\end{equation}

Presently, we have measured the magnetic field strengths in
different lines. In the following sections we will use the LOS
magnetic field, but not the strength, for two reasons. First, the
field in the region is mainly vertical and the region is not so
far from the solar disk center. Secondly, only the LOS magnetic
field can be computed for the Na D$_1$ line. In the flux tube
model the tube expands and the magnetic field strength decreases with
height. Equation (\ref{eqn13}) would indicate that the formation
height of Fe {\sc \romannumeral 1} 5250.2 \AA \ is the highest one,
and the Ca {\sc \romannumeral 1} 6102.7 \AA \ is formed below the
Fe {\sc \romannumeral 1} 6302.5 \AA.

We can also test this assumption of the flux tube model by measuring
the full-width at half-maximum of the absolute value of the field
strength variation along a cut in the image. We analyzed the LOS
magnetic fields along the selected cuts presented in Figure
\ref{fig02}. They are plotted in the top row of Figure
\ref{fig08}. The full-widths at half-maximum follow the relations:
\begin{equation}
W_\mathrm{LOS,6102.7} > W_\mathrm{LOS,6302.5} >
W_\mathrm{LOS,5250.2} \ .\label{eqn14}
\end{equation}
In the flux tube model we expect that the structure with the
smallest width would be formed at the lowest height. Equation
(\ref{eqn14}) is in contradiction with what Equation (\ref{eqn13})
implies. We can use the measurements of the magnetic field along
the Na D$_1$ line wings, where the relative formation heights of
each wavelength have already become known, to explain this
inconsistency.

\subsection{Magnetic Field Gradient in the Na D$_1$ Line Wings}

\begin{figure}
\centerline{\includegraphics[width=1\textwidth]{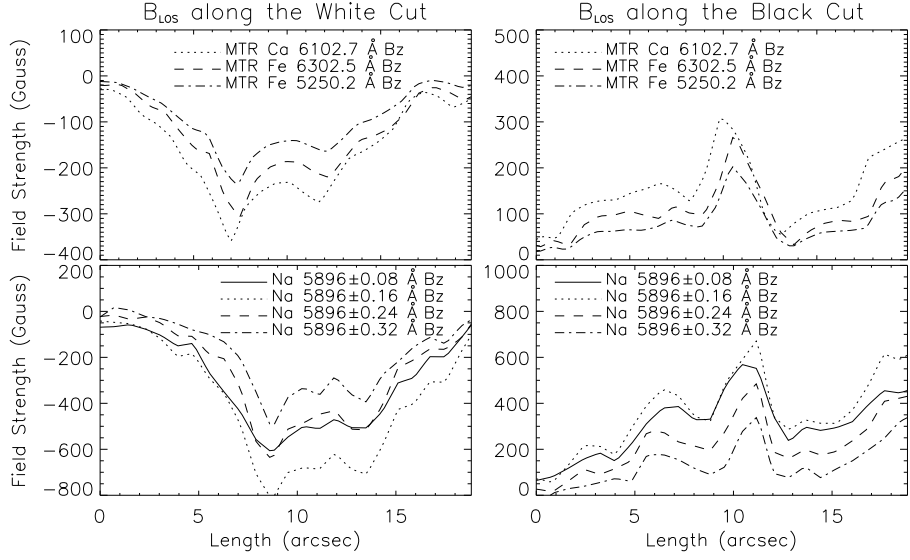}}
\caption{LOS magnetic fields along the white (left column) and
black (right column) cuts as shown in Figure \ref{fig02}. Top row
shows the distributions for Ca {\sc \romannumeral 1} 6102.7 \AA \
(dotted), Fe {\sc \romannumeral 1} 6302.5 \AA \ (dashed), and Fe
{\sc \romannumeral 1} 5250.2 \AA \ (dash-dotted). Bottom row shows
for Na D$_1$ 5896 \AA \ with $\Delta\lambda=$ 0.08 \AA \ (solid),
0.16 \AA \ (dotted), 0.24 \AA \ (dashed), and 0.32 \AA \
(dash-dotted).}\label{fig08}
\end{figure}

We have computed the fields at different wavelengths in the line
wing of Na D$_1$ 5896 \AA, {\it i.e.} $\Delta\lambda=$ 0.08 \AA,
0.16 \AA, 0.24 \AA, and 0.32 \AA. The LOS magnetic fields along
the white and black cuts as shown in Figure \ref{fig02} are
plotted at the bottom row of Figure \ref{fig08}. We measured the
full-widths at half-maximum of the structures, which peak at 6 to
12 arcsec in all the plots of Figure \ref{fig08}, and we found the
following relation:
\begin{equation}
W_\mathrm{LOS,5896\pm0.08} > W_\mathrm{LOS,5896\pm0.32} \ .\label{eqn15}
\end{equation}
The magnetic features observed in the line core have a larger size
than that in the line wing. This result is consistent with the
flux tube model assumption. Let us analyze the magnetic field
strength variation along the Na D$_1$ profile.

From the bottom row in Figure \ref{fig08}, we got the following
relation of the magnetic fields obtained in the core and wing of
Na D$_1$ ($\Delta\lambda=$ 0.08 \AA \ and 0.32 \AA \
respectively):
\begin{equation}
|B_\mathrm{LOS,5896\pm0.08}| > |B_\mathrm{LOS,5896\pm0.32}| \ , \label{eqn16}
\end{equation}
The fields in the line core of Na D$_1$ are stronger than those in
the line wing. We found a similar contradiction by using Ca {\sc
\romannumeral 1} and Fe {\sc \romannumeral 1} lines. Such a behavior
for the Na D$_1$ line has been reported by \inlinecite{Berlicki2006}
and \inlinecite{Mein2007}.

\section{Discussion}\label{s-magalt}

\subsection{Magnetic Flux Tube Modelling}

\begin{figure}
\centerline{\includegraphics[width=0.5\textwidth,angle=-90,viewport=150
100 447 694, clip]{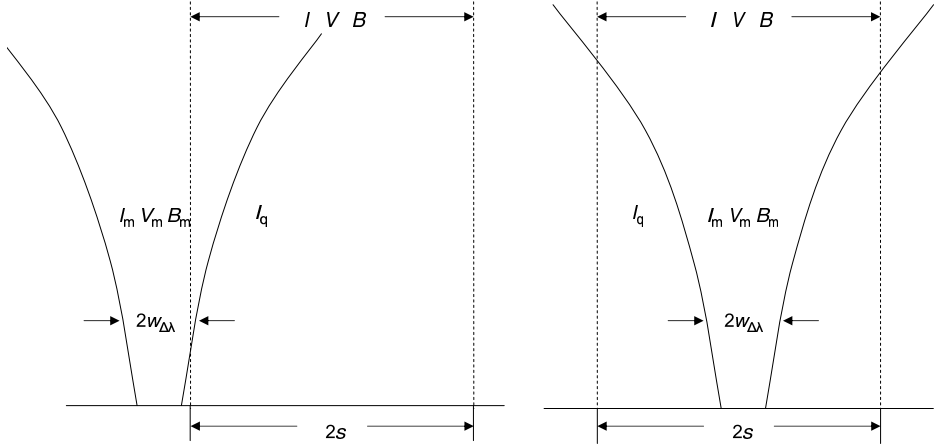}} \caption{The sketch of a flux tube
model and the parameters for the flux tube. $I_\mathrm{m}$,
$V_\mathrm{m}$, and $B_\mathrm{m}$ are the Stokes parameters and
LOS magnetic field at the flux tube center. $I$, $V$, and $B$ are
the observed parameters convolved with the quiet sun region. The
intensity of the quiet sun is $I_\mathrm{q}$. The half-widths of
the tube and the spatial resolution are denoted by
$\mathrm{w}_{\Delta \lambda}$ and $s$, respectively. All the
parameters except $s$ depend on the wavelength position $\Delta
\lambda$ from the line center of Na D$_1$ 5896 \AA. Left and right
panels show observations at the edge and at the center of the flux
tubes, respectively.}\label{fig09}
\end{figure}

\inlinecite{Mein2007} resolved this contradiction by a 2D magnetic
flux tube model. They computed theoretical profiles of the Na
D$_1$ line in an atmosphere where flux tubes and quiet sun regions
coexist side-by-side in a pixel. First, they showed that the
increase of magnetic field strength with height can be explained
by a pure geometrical effect at the edge of flux tubes. For one
pixel with a size of 2$s$, where $s$ is the half-width of the
spatial resolution, a smaller quiet sun region should be
considered at higher levels than at lower altitudes ({\it left
panel} of Figure \ref{fig09}). Secondly, it should be explained by
the combination of small filling factors with slopes of the Stokes
profiles at the flux tube center ({\it right panel} of Figure
\ref{fig09}). In fact, the measurements of Stokes parameters and
magnetic field ($I$, $V$, and $B$) are obtained by THEMIS with a
spatial resolution, 2$s$, much larger than the expected size of the
flux tube sections 2$\mathrm{w}_{\Delta \lambda}$.
\inlinecite{Mein2007} studied this effect on the Stokes profiles
by using flux tube models with different sizes. On the one hand,
the filling factor of the lower formed line is smaller because the
size of the flux tube is smaller. On the other hand, the Stokes
$I$ profile slope depends on the atmosphere model. In the line
wing, the absolute value of the profile slope $\frac{\partial
I}{\partial \lambda}$ is larger in the quiet sun than that in the
flux tube center. It is reversed in the line core. Assuming that
$I_\mathrm{q}(\Delta \lambda)$ and $I_\mathrm{m}(\Delta
\lambda,0)$ are the intensities in the quiet sun region and
magnetic flux tube center as denoted in Figure \ref{fig09}, we
have the following relationship for $|\Delta \lambda| >$0.24 \AA:
\begin{equation}
|\frac{\partial I_\mathrm{q}(\Delta \lambda)}{\partial \lambda}| > |\frac{\partial I_\mathrm{m}(\Delta \lambda,0)}{\partial \lambda}| \ ,
\end{equation}
and for $\Delta \lambda <$0.24 \AA:
\begin{equation}
|\frac{\partial I_\mathrm{q}(\Delta \lambda)}{\partial \lambda}| < |\frac{\partial I_\mathrm{m}(\Delta \lambda,0)}{\partial \lambda}| \ .
\end{equation}
The ratio $r(\Delta \lambda,0)$ of the profile slope between the
quiet sun and flux tube region is larger in the line wing than in
the line core, where
\begin{equation}
r(\Delta \lambda,0) = \frac{\partial I_\mathrm{q}(\Delta \lambda)}{\partial \lambda}/\frac{\partial I_\mathrm{m}(\Delta \lambda,0)}{\partial \lambda} \ .
\end{equation}
\inlinecite{Mein2007} derived the following equation from a
simpler but comprehensive model, where the magnetic field in the
flux tube is constant at the same height:
\begin{equation}
\frac{B_\mathrm{m}(\Delta \lambda,0)}{B(\Delta \lambda,0)} = 1 + r(\Delta \lambda,0)\cdot(s/\mathrm{w}_{\Delta \lambda} - 1) \ .\label{eqn110}
\end{equation}
The half-widths of the tube and the spatial resolution of the
instrument are denoted by $\mathrm{w}_{\Delta \lambda}$ and $s$,
respectively, as shown in Figure \ref{fig09}. $B_\mathrm{m}(\Delta
\lambda,0)$ and $B(\Delta \lambda,0)$ are the LOS magnetic field
in the flux tube center and the synthetic magnetic field convolved
with the quiet sun region within the spatial resolution $2s$,
respectively. $B(\Delta \lambda,0)$ corresponds to the observed
magnetic field.

$B(\Delta \lambda,0)$ has a negative relationship with the ratio
$r(\Delta \lambda,0)$, which is larger in the line wing, and a
positive relationship with $\mathrm{w}_{\Delta \lambda}$, which is
smaller in the line wing. So $B(\Delta \lambda,0)$ will become
smaller at lower altitudes after convolving the profiles of the
flux tubes with the profiles originating from the quiet sun.

The discussion above is for a single flux tube, whose size may be
smaller than the spatial resolution of SOT/SP ($0.3''$). In this
case, each of the peaks $A1, A2, B1, B2$, and $B3$ measured by
SOT/SP shown in Figure \ref{fig06} is constituted by a bundle of
smaller flux tubes. \inlinecite{Mein2007} showed that the sizes of
the flux tube bundle still increase with height in a multi-flux
tube simulation. Further, Equation (10) still holds, {\it i.e.} a
combination of Stokes profiles in quiet and active regions may
reverse the field strength variation with height.

\subsection{Relative Formation Heights of the Fe {\sc \romannumeral 1}, Ca {\sc \romannumeral 1}, and Na D$_1$ Lines}
The results of Equation (\ref{eqn16}), in contradiction
with the half-width relation, are now explained by the small
filling factor of the magnetic field and the difference of slopes
of the Stokes $I$ profiles observed in quiet sun and in the flux
tube. We believe that these effects will also affect the Ca {\sc
\romannumeral 1} and Fe {\sc \romannumeral 1} lines. Thus, we
suggest that the formation altitude of Ca {\sc \romannumeral 1}
6102.7 \AA \ is higher than that of Fe {\sc \romannumeral 1}
6302.5 \AA, which forms higher than Fe {\sc \romannumeral 1}
5250.2 \AA. The formation height of Ca {\sc \romannumeral 1}
6102.7 \AA \ is comparable to the one where the line wing of Na
D$_1$ forms. Nevertheless, we should still keep in mind that the
magnetic fields obtained by Ca {\sc \romannumeral 1} 6102.7 \AA \
were computed in LTE and the Milne--Eddington approximation, which
may be not adequate for this line. The results for the two Fe {\sc
\romannumeral 1} lines are more reliable; however, the formation
height of a line varies along the line profile itself. In
addition, the magnetic inversion makes use of the whole profile,
which forms at different altitudes. Nevertheless, if Fe {\sc
\romannumeral 1} 6302.5 \AA \ forms at greater heights than Fe
{\sc \romannumeral 1} 5250.2 \AA \ at each wavelength from the
line center, this relation will be conserved by the inversion.

\section{Conclusion}\label{s-discu}

In this paper we have compared the vector magnetic fields obtained
from the THEMIS/MTR multi-line observations and {\it Hinode}
SOT/SP observations for a facular region. Two inversion codes,
{\it i.e.} UNNOFIT and MELANIE, were adopted to determine the
field strength, azimuth, and inclination angles in this region
from the full Stokes profiles. The 180$^\circ$ ambiguity was
removed by the non-potential magnetic field calculation method
\cite{Georgoulis2005,Metcalf2006}. The magnetic fields in the
faculae are found to be mainly vertical. We resolve small flux
tube bundles in the faculae with converging and diverging
horizontal components of magnetic fields in negative and positive
polarities, respectively. The maximum field strengths observed by
SOT/SP (1200 gauss) are four times larger than those obtained from
MTR. The sizes of the detectable facular details are found to be
about $1''$ to $2''$ for SOT/SP, while they are larger when
measured by MTR. Nevertheless, we conclude that MTR and SOT/SP
give consistent results concerning the characteristics of the
fields, particularly for the inclination and azimuth histograms.
The differences between the field strengths are reduced to a
factor of $\approx 1.5$ if the Earth's atmospheric seeing effects
are taken into account.

The assumption of the conservation of magnetic flux in the
expanding flux tube model has been used to derive the formation
heights of the lines observed with MTR. We also consider the
opposite effect on the magnetic field strength gradient that is
generated by poor spatial resolution compared to the size of the
flux tube \cite{Mein2007}. Fe {\sc \romannumeral 1} 6302.5 \AA \
forming at greater heights than Fe {\sc \romannumeral 1} 5250.2
\AA \ can be compared with some simulated results listed in Table
\ref{tbl13}. It is consistent with the calculation by
\inlinecite{Bommier2007} obtained by solving the radiative
transfer equation.

The combination of the high spatial resolution of SOT/SP and the
multi-line observation of THEMIS/MTR allowed us to have a 3D view
on the flux tubes in faculae. To get a quantitative model, one
would need to calculate more complete and reliable formation
heights of the lines using response functions. The formation
height of different magnetically sensitive lines is an important
parameter in the new 3D method to resolve the 180$^\circ$
ambiguity, where the divergence-free nature of magnetic fields is
used. Removing the 180$^\circ$ ambiguity is
necessary to built vector magnetic fields, which are used as the
boundary condition for nonlinear force-free field extrapolations
to get the fields in the corona.


%

\begin{table}
\caption{Instrumental characteristics for different lines of
THEMIS/MTR and \emph{Hinode} SOT/SP. The data were observed on 23
May 2008.}\label{tbl11}
\begin{tabular}{c c c c}
\hline
Line & Scan duration & Dispersion & Pixel size \\
 & (UT) & (\AA /pixel) & (arcsec) \\
\hline
\multicolumn{1}{r}{MTR Ca {\sc \romannumeral 1} 6102.7 \AA} & 16:44:03 -- 17:38:55 & 0.0130 & 0.800, 0.236 \\
\multicolumn{1}{r}{MTR Fe {\sc \romannumeral 1} 6302.5 \AA} & 16:44:03 -- 17:38:55 & 0.0125 & 0.800, 0.233 \\
\multicolumn{1}{r}{MTR Fe {\sc \romannumeral 1} 5250.2 \AA} & 16:44:03 -- 17:38:55 & 0.0111 & 0.800, 0.230 \\
\multicolumn{1}{r}{MTR Na D$_1$ 5895.9 \AA} & 16:44:03 -- 17:38:55 & 0.0130 & 0.800, 0.232 \\
\multicolumn{1}{r}{MTR H$\alpha$ 6562.8 \AA} & 16:44:03 -- 17:38:55 & 0.0144 & 0.800, 0.226 \\
\multicolumn{1}{r}{SOT/SP Fe {\sc \romannumeral 1} 6302.5 \AA} & 14:30:05 -- 15:01:38 & 0.0215 & 0.297, 0.320 \\
\hline
\end{tabular}
\end{table}

\begin{table}
\caption{The free parameters fitted in UNNOFIT and MELANIE.}\label{tbl12}
\begin{tabular}{r r}
\hline
UNNOFIT & MELANIE \\
\hline
Zeeman splitting $\Delta\lambda_{\mathrm H}$ & Field strength $B$ \\
Inclination angle $\psi$ & Inclination angle $\psi$ \\
Azimuth angle $\phi$ & Azimuth angle $\phi$ \\
Line strength $\eta_0$ & Line strength $\eta_0$ \\
Doppler width $\Delta\lambda_{\mathrm D}$ & Doppler width $\Delta\lambda_{\mathrm D}$ \\
Damping parameter $\gamma$ & Damping constant $a$ \\
Line wavelength $\lambda_0$ & Doppler velocity $v$ \\
Parameter of source function $b$ & Source function gradient $B_1$ \\
Filling factor $\alpha$ & Stray light fraction $f$ \\
 & Macro-turbulence velocity $v_{\mathrm m}$ \\
\hline
\end{tabular}
\end{table}

\begin{table}
\caption{Formation heights at the line center of Fe {\sc
\romannumeral 1} 6302.5 \AA \ and Fe {\sc \romannumeral 1} 5250.2
\AA. The heights are measured from $\tau_{5000} = 1$.}
\label{tbl13}
\begin{tabular}{c  c  c  c}
\hline Method & \multicolumn{2}{c}{Formation height (km)} & Atmosphere model \\
 & Fe {\sc \romannumeral 1} 6302.5 \AA & Fe {\sc \romannumeral 1} 5250.2
 \AA & \\
\hline
Bommier {\it et al.} (2007) & 262 & 210 & Maltby {\it et al.} (1986) \\
Bruls {\it et al.} (1991) & 232 & 259 & VAL-C \\
Sheminova (1998) & 404 & 440 & Holweger-M\"uller
 \\
\hline
\end{tabular}
\end{table}

%

\begin{acks}
THEMIS is a French--Italian telescope operated by the CNRS and CNR
on the island of Tenerife in the Spanish Observatorio del Teide of
the Instituto de Astrof{\'i}sica de Canarias. {\it Hinode} is a
Japanese mission developed and launched by ISAS/JAXA, with NAOJ as
domestic partner and NASA and STFC (UK) as international partners.
It is operated by these agencies in cooperation with ESA and NSC
(Norway). The authors thank P. Mein, M.K. Georgoulis, N. Vitas,
and T. T$\mathrm{\ddot{o}r\ddot{o}}$k very much for their help and
discussions. Y. Guo was supported by the scholarship granted by
China Scholarship Council (CSC) under file No. 2008619058. S.
Gosain acknowledges CEFIPRA funding for his visit to Observatoire
de Paris, Meudon, France under its project No. 3704-1.
\end{acks}

 \bibliographystyle{spr-mp-sola-cnd} 
 \bibliography{bibliography_sci}

\end{article}
\end{document}